\documentclass[3p, preprint, times]{elsarticle}
\graphicspath{ {./figures/} }
\usepackage{natbib}
\usepackage{hyperref}
\usepackage{float}
\usepackage{amsmath}
\usepackage{amssymb}
\usepackage{verbatim} %comments
\usepackage{apalike}
\usepackage{mathrsfs,amsmath}
\restylefloat{figure}
\restylefloat{table}
\hypersetup{colorlinks}
\usepackage{color}
\usepackage{array}
\usepackage{multirow}
\usepackage{algpseudocode}
\usepackage{algorithm}
\biboptions{authoryear}

\begin{document}
\begin{frontmatter}

\title{GCNET: A novel graph-based approach for prediction of stock price movement}

\author[label1]{Alireza Jafari}
\ead{alireza.jafari7@ut.ac.ir}

\author[label1]{Saman Haratizadeh\corref{cor1}}
\ead{haratizadeh@ut.ac.ir}

\cortext[cor1]{Corresponding author}
\address[label1]{Faculty of New Sciences and Technologies, University of Tehran, North Kargar Street, 1439957131 Tehran, Iran}

\begin{abstract}

The importance of considering related stocks data for the prediction of stock price movement has been shown in many studies, however, advanced graphical techniques for modeling, embedding and analyzing the behavior of inter-related stocks have not been widely exploited for the prediction of stocks price movements yet. The main challenges in this domain are to find a way for modeling the existing relations among an arbitrary set of stocks and to exploit such a model for improving the prediction performance for those stocks. The most of existing methods in this domain rely on basic graph-analysis techniques, with limited prediction power, and suffer from a lack of generality and flexibility. In this paper, we introduce a novel framework, called GCNET that models the relations among an arbitrary set of stocks as a graph structure called influence network and uses a set of history-based prediction models to infer plausible initial labels for a subset of the stock nodes in the graph. Finally, GCNET uses the Graph Convolutional Network algorithm to analyze this partially labeled graph and predicts the next price direction of movement for each stock in the graph.

GCNET is a general prediction framework that can be applied for the prediction of the price fluctuations of interacting stocks based on their historical data. Our experiments and evaluations on a set of stocks from the NASDAQ index demonstrate that GCNET significantly improves the performance of SOTA in terms of accuracy and MCC measures.

\end{abstract}


\end{frontmatter}

\section{Introduction}
\label{introduction}

Predicting the stock prices and fluctuations of stock prices has been of great interest for decades since any reliable information about the possible next price or direction of price movement can be of great value for investors who need to decide how to invest in the market  \citep{rather2017stock,soni2011applications}. Traditional stock prediction approaches are categorized into technical analysis and fundamental analysis. While fundamental analysis focuses on the intrinsic value of stocks and is mainly used by human experts, in technical analysis, the history of stock prices and price indicators are analyzed to extract useful patterns for the prediction of the future behavior of the stocks prices. 

In past decades, technical analysis has been widely used to build automatic stock market prediction systems \citep{bustos2020stock}. Researchers have applied machine learning techniques to develop a wide variety of prediction models, including regression and classification systems, for forecasting the future prices or price fluctuations of stocks. Especially during the past few years, along with the advancements in the deep learning domain, sophisticated deep models have been developed for market forecasting, outperforming their traditional predecessors \citep{ hoseinzade2019cnnpred, li2017comparative, patel2015predicting, gunduz2017intraday}. 
These experiments show that deep methods can model the complex behavior of markets more accurately. However, in most of these studies, the prediction models use the historical data from the past prices of a single stock to predict the future of that stock's price and ignore the relationships between stocks as a source of information. 

One emerging idea in this domain is that financial entities like markets or stocks are naturally connected entities having interactions with each other, and so, it is reasonable to believe that the behavior of one entity can affect the others. This means that intelligent analysis and consideration of related entities can help us to predict the future behavior of others.

Following this idea, some studies have used historical data from other possibly related stocks/markets as well to train prediction models for a target stock/market \citep{hoseinzade2019cnnpred, zhong2017forecasting, gunduz2017intraday}. Although sophisticated deep forecasting models have been used in this class of studies, they still have some limitations: They usually rely on an expert's advice to select a preferably small set of related stocks/markets to the target entity. Also, most of the existing models in this category are not flexible enough to explicitly model and use the direct and indirect relationships among entities or to reflect the already known degrees of relevance between them. Another drawback of this approach is that the architectures of these models are usually task-dependent and are designed for prediction over a fixed set of related entities and they may not be easily applicable for other, possibly larger, sets of entities. 

A natural approach to model the inter-relations among a set of entities is to use graphical analysis that can discover the hidden relationship between stocks from the graph structure, unlike previous works, which use the manual selection of related stocks. Graph-based modeling and analysis of data is a standard technique in many domains such as social network analysis or recommendation systems, however, it has not been widely used for modeling relations among financial entities and its application in this domain has been limited to a few studies \citep{yin2022graph, kia2020network, long2020integrated, kia2018hybrid, kim2017predicting}.

The common approach in these studies is to construct a (usually correlation-based) graph structure reflecting some kind of relation among entities and then propagate a set of already known labels to other nodes through the paths that exist among nodes in the graph. Although these algorithms have achieved some improvement in performance over history-based prediction techniques, they still suffer from some important flaws. Firstly, they rely on a set of known labels for some entities while that may not be available in all settings. Secondly, their prediction mechanism is solely based on the structure of the graph and usually ignores any local information about each entity's current state for prediction of its. Finally, these algorithms in their prediction step usually use neighbor-based approaches that cannot be trained or optimized based on the current state of a node and its neighbors or the structure of the current underlying graph. This lack of adaptability makes the performance of the prediction step extremely dependent on the initial structure of the graph, which limits the prediction performance especially when there are hidden interactions among entities that are not reflected in the graphical model. 

Fortunately, in recent years there have been significant developments in the field of graph analysis using Graph Neural Networks. These techniques are able to analyze graph-structured data in order to generate new representations of each node while considering node features. In this paper, we introduce GCNET, a general framework for the prediction of stock price fluctuations that formulates this classic problem as a label prediction task in a graph. It models the price historical data of a set of inter-related stocks (or markets) along with the observed degrees of co-relations among them as a graph and creates beneficial aggregation for each node by considering its related neighbor stocks. GCNET uses a deep graphical semi-supervised learning algorithm to generate a model for predicting the next price (or index) fluctuation for each stock (or market).

The main characteristics and contributions of the introduced framework, GCNET, can be summarized as follows:
\begin{itemize}
    
    \item{We introduce a novel framework to formulate the prediction of stock price fluctuations as a label prediction task over a  graphical model of inter-related financial entities.}
    \item{Our suggested framework can exploit any set of history-based predictors to generate the initial labels required by its semi-supervised learning process and so it is applicable in settings where there is no already known label/class for any node/stock.}
    \item {We introduce a novel approach for modeling the relations among stocks as a so-called influence network that improves its overall prediction performance.}
    \item{A comprehensive set of experiments on two sets of hundred stocks shows that GCNET predicts the next price fluctuations for the stocks in the graph with a significantly better performance compared to the state of the art baselines.}
\end{itemize}

In the next section, we will review the related work. Our network construction method and prediction algorithm are explained in Section \ref{model}. Section \ref{experiment} presents detailed information about the dataset used in our experiments, the experimental setup, the results, and a discussion about the observations. Finally, we conclude the paper in Section \ref{conclusion}.

\section{Related Works}
\label{related_work}

Many different machine learning techniques have been used for the prediction of stock markets in the past decades. Most of these techniques rely on analyzing the historical data of a single stock data to extract useful patterns for predicting its future behavior. \cite{kara2011predicting} used ANN and SVM to predict the direction of movement of the Istanbul Stock Exchange (ISE) National 100 Index and showed that ANN predicts the index significantly better than SVM. \cite{zhong2017forecasting} used ANN to predict S\&P500 in their model and showed that using PCA to construct new representations for the initial feature vectors can improve the prediction accuracy of the ANN-based model. \cite{arevalo2016high} examined artificial neural networks with different structures and observed that a deep neural network with five hidden layers achieves the highest accuracy in predicting the direction of movement for a single stock's price. Long Short-Term Memory (LSTM) is one of the most popular models used for time series prediction. \cite{nelson2017stock} used LSTM and MLP for forecasting the Brazilian stock market. The results showed the superiority of LSTM over MLP. In a more recent study, LSTM has been combined with an attention mechanism to predict the market shares based on the price information from the past few days \citep{feng2019enhancing}. Although only one stock data is used in this model to predict, in more recent studies, researchers have emphasized the importance of considering related stocks by trying to use other stocks in their model training using simple techniques. \cite{shah2021stock} analyzed historical data of different stocks by Bi-directional Long Short-Term Memory and predict stock price trends. They also presented a framework for depth and time calculation learning faster than the one-directional approach.

Convolution neural network, CNN, is another class of deep learning algorithms used in stock market forecasting while its ability to extract high-level features from price data has been studied by several researchers \citep{gunduz2017intraday, hoseinzade2019cnnpred, di2016artificial}. In one study, \cite{di2016artificial} different kinds of deep neural networks including ANN, LSTM, and CNN were used to predict the S\&P500 index, and the results showed that CNN outperforms other deep models in prediction.
\cite{gunduz2017intraday} used CNN to predict the market index of the Istanbul Stock Exchange.They used a handmade set of stocks to input their model to consider related stocks. \cite{hoseinzade2019cnnpred} predicted well-known stock market indices in the United States using CNN. They showed that 2D and 3D CNNs could be applied to combine information from a fixed set of related time series in order to successfully predict their future behavior. Although the above models have yielded acceptable results, most of them do not use stock relationships as a valuable source of information and the rest are limited to using the handful of stocks that the authors guess are relevant.

Another class of algorithms for financial time series prediction are the Graph-based methods \cite{yin2022graph}, \cite{kia2020network}, \cite{long2020integrated}, \cite{kia2018hybrid}, \cite{kim2017predicting}, and \cite{shin2013prediction}. 
The majority of these methods model the relations of the entities as correlation graphs and use models to predict that can not be easily generalized to other sets of stocks or markets.
\cite{park2013stock} created a network of nodes representing the stocks in the Korean Stock Exchange and some commodity price time series based on the similarity of their feature vectors. They assigned the known labels based on the amount of variation in the 5-day moving average of stock prices and predicted the direction of movement for the Korean Stock Exchange index using a label propagation algorithm. 
\cite{kim2017predicting} used network analysis and showed that graph structures can reveal important information about the future of stock prices. Taking into account the parameters of the network, they developed an ARIMA model and predicted the value of the S\&p500 index.
In \cite{kia2018hybrid} a graph structure is introduced in which the nodes represent the markets, and the edges represent strong correlations between the market index time series. The nodes whose corresponding indices are already known, are labeled while the nodes whose corresponding indices are still unknown, due to the time differences between different time zones, are predicted using a label spreading algorithm.
\cite{long2020integrated} creates a knowledge graph representing different companies and apply the node2vec algorithm to select the relevant stocks of the target for constructing helpful embedding \citep{grover2016node2vec}. They use the similarities between the resulted embeddings as a quantitative measure of the co-relation between stocks. They weight each stock using these co-relation values and train a LSTM model for prediction without exploiting the power of network analysis in the forecasting process.

In \cite{kia2020network} an association rule mining technique is used for defining the weights of edges in a graph of markets, and to predict the unknown labels in the graph, a variation of PageRank algorithm is used. As we mentioned before, the simple label prediction algorithms used in this class of methods, not only rely on the existence of already known labels but also they fail to model possibly complex relations among the internal states of the nodes, their inter-relations, and their labels. Also, despite the fact that the performance of these algorithms is highly dependent on the structure of the underlying graph, the graph construction process needs to be subject to further studies in this domain.
Like most of the mentioned models, \cite{yin2022graph} uses the correlation graph to predict the stock price of the Chinese stock market. They create new representations by attention mechanism and then use these vectors to train and predict the LSTM model.
This class of models relies entirely on underlying graph structures, that are mostly defined based on historical price correlations among stocks, while other possible graph structures have received less attention.

As a modern node embedding and label prediction algorithm, Graph Convolutional Network, GCN, has been recently introduced. GCN is an extension to CNN that is able to handle graphical data \cite{kipf2016semi}. It is a Graph Neural Network (GNN) that takes the graph structure and node features as inputs and aggregates and transforms information from neighbors of each node to create a new representation for that node which is more informative for the prediction of a target variable like the label or class of the node. In this context, nodes can be stocks and edges indicate the mutual importance of stocks. GCN can construct representations of related stocks by analyzing the underlying graph while considering both stock relationships and the initial feature vector of each stock. GCN has recently received extensive attention because of its outstanding performance on node classification tasks in various fields, including recommender systems \cite{hekmatfar2021embedding}, traffic prediction \cite{zhao2019t} and, cancer diagnosis \cite{zhou2019cgc}. In the field of stock market prediction, GNNs are limited to a few studies due to their recent emergence \citep{kim2019hats, hu2018listening}. \cite{kim2019hats} used a graph attention network to forecast stocks. For creating their graph, they used a classic dataset of Wikidata in previous work \citep{vrandevcic2014wikidata}, and their infrastructure network was very simple. GNN has been used in a few studies for representational learning from textual data available in social media, news articles, and blogs \citep{hu2018listening} but they have not been widely applied in the field of graph-based financial prediction to analyze financial graphical models yet.

In the next section, we will introduce GCENT, as a novel framework for graph-based prediction of the direction of stocks price movements. GCNET resolves the shortcomings of the existing methods by introducing a novel approach to modeling the prediction of the price fluctuations for a set of inter-connected stocks in a market as a semi-supervised label prediction process using GCN. It also presents a new technique for constructing the graphical model of the data and introduces an effective method for labeling a subset of the graph nodes using reliable history-based predictions as well.

\section{Model description}
\label{model}
In this section, we are going to explain our suggested semi-supervised prediction algorithm called GCNET. As summarized in Algorithm \textcolor{mycolor}{1}, GCNET first models the stock market as a complex network structure of the stocks' historical data. Using the structure of the created graph, for each day of the stock market separately, we create a graph of nodes, which represent stocks and we have introduced a novel technique for connecting and weighting graph edges. Also, each node contains a feature vector of technical indicators from the specified day of corresponding stock.
In the next step, we assign a set of initial labels to a portion of the nodes of test day using a set of reliable predictions made by a so-called PLD method. Each label in this step represents a first guess about the possible next price fluctuation for the corresponding stock.
The algorithm then uses a GCN technique to process the resulted partially labeled graph, in order to refine the initial labels and also predict the labels for the unlabeled subset of the nodes. The final labels form the algorithm's predictions for the next-day fluctuations of the stocks. In the following subsections, we explain the details of the GCNET steps. 

\begin{algorithm}[H]

\hspace*{\algorithmicindent}
\textbf {Input :}\\
\hspace*{\algorithmicindent}\hspace*{\algorithmicindent}\textbf{Dataset} (Price history for m stocks) \\ 
\hspace*{\algorithmicindent}\textbf{Output :}\\
\hspace*{\algorithmicindent}\hspace*{\algorithmicindent}\textbf{List of predicted labels}
\begin{algorithmic}
\vspace{-3mm}
\\
\hrulefill
\State $G \gets$ Generate\_Influence\_Graph(Dataset) 

\State $G \gets$ PLD(Dataset,G) \Comment{label a subset of the nodes}
\State $G \gets$ Add node feature vectors to G
\State $Predictor\gets$Train GCN on G
\State $L\gets Predictor(G)$ \Comment {predict labels using the trained GCN}
\State return (L)

\caption{GCNET algorithm}
\end{algorithmic}
\label{alg:GCNET}
\end{algorithm}

\subsection{Network Construction}
\label{infnet}

In this section, we introduce a graph generation algorithm for constructing a novel so-called influence network, which is designed to serve the needs of the subsequent label prediction mechanism used in the GCNET framework. This algorithm tries to generate a network containing paths through which sharing information among nodes may improve the performance of the target prediction task for each node. In the following sub-sections, we explain the details of the graph structure and the algorithm used for generating it.

In our graph the nodes represent stocks. If there are $m$ stocks in the dataset: $ Stocks = \lbrace s_{1} , s_{2} , ..., s_{m} \rbrace $, we denote the graph as $G = (V , E)$, in which $V = \lbrace v_{s_1} , v_{s_2} , ..., v_{s_m} \rbrace $ is the set of nodes and the set $E$ is the set of edges. Each edge $e_{i,j}$ connects the pair of nodes $v_{s_i}$ and $v_{s_j}$, where $i \neq j$. The edges are weighted and we denote the weight of the edge $e_{i,j}$ as $w_{i,j}$.

%\subsubsection{Influence network}
%\label{infnet}

To add edges to the graph, GCNET first calculates an influence score for each pair of nodes. This score estimates how useful is to combine the historical information from those stocks for the prediction of each one's future fluctuation. If this aggregation of information has a positive influence on the prediction performance, then an edge connecting that pair of nodes is added to the graph, with a weight reflecting the amount of that influence. 

GCNET needs to build several simple prediction models to construct the influence graph with n nodes: For each target stock, one model is trained using that stock's historical data and $n-1$ models are trained, each using the historical data from the target stock and one other stock in the graph. To achieve a reasonable computation time for training the required prediction models we use Quadratic Discriminant Analysis that is a fast and effective history based learning algorithm. Quadratic discriminant analysis (QDA) is a well-known statistical classification technique that is a variant of linear discriminant analysis that allows for non-linear separation of data. QDA classifier is attractive because it has closed-form solutions that can be easily computed, work well in practice, and have no hyper-parameters to tune. QDA finds an efficient decision rule when a linear discriminant procedure is not sufficient to separate the groups under study or when the covariance matrices are not equal among groups \citep{james2013introduction}.

QDA assumes multivariate data, following a normal distribution, and uses the Bayes theorem to calculate the probability a sample $x$ belongs to class $k$. However, a separate covariance matrix $\Sigma_k$, is assumed for each class, $k = 1, 2, ..., k_n$ ($n$ = number of classes), yielding the quadratic discriminant function as:

\begin{equation}
	\delta_k(x) = -\frac{1}{2} log|\Sigma_k| -\frac{1}{2}(x-\mu_k)^T \Sigma_k^{-1} (x-\mu_k) + log \pi_k
\end{equation}

where $\delta_k$ is the score value of discriminant for class $k$, $\Sigma_k$ is the covariance matrix and $\pi_k$ is the prior probability of the k-th population. Class $Clf$, that sample x belongs to is then predicted as:
\begin{equation}
	  Clf(x) =  \underset{k}{\text{arg max }} \delta_k(x)
\end{equation}

To train the models, for each stock $s_i$ a feature vector $v_i$ is extracted from its historical data based on the set of features summarized in Table 1. Also, for each pair of stock  $s_i$ and $s_j$ an average feature vector is defined as $v_{ij}=v_{ji}=average(v_i,v_j)$. Then GCNET uses QDA to train four prediction models $P_i$, $P_j$, $P_{ij}$ and $P_{ji}$ for each pair $s_i,s_j$ of stocks. $P_i$ denotes the model predicting the next price movement for stock $S_i$ using $v_i$ as input, while $P_{ij}$ uses $v_{ij}$ as the input of the model to predict the next price movement for stock i ( Similar definitions hold for $P_j$, $P_{ji}$). 

Suppose that $Acc_i$, $Acc_{ij}$, denote the prediction accuracy of $P_i$, $P_{ij}$ for stock $i$ while $Acc_j$ and $Acc_{ji}$ represent the accuracy of $P_j$, $P_{ji}$ in prediction of the next price movement for $s_j$ (all measured on the validation data). The $influence_{\{i,j\}}$ is then defined as follows: 

\begin{align}
\label{weight_influ}
	\begin{split}
		& influence_{\{i,j\}} = \frac{(Acc_{ij} - Acc_i)+(Acc_{ji}-Acc_j)}{2} \\
	\end{split}
\end{align}

Clearly, for a pair of stocks $s_i$ and $s_j$, $influence_{\{i,j\}}$ reflects the average improvement in prediction performance achieved for each stock by using the other stock's information as well in the prediction model. Based on this measure then the weight of the edge connecting the corresponding nodes in the graph is calculated as follows:

\begin{align}
\label{weight_edge}
	\begin{split}
		& w_{i,j}=w_{j,i}= \max (influence_{\{i,j\}},0)
	\end{split}
\end{align}

In other words, the weight of the edge between two stock nodes represents how much the aggregation of information between those two nodes we expect to boost the performance of the resulting predictions for each stock.
The intuition of using this technique in GCNET comes from the behavior of the aggregation step in the GNN algorithms such as GCN. Since this graph is going to be ultimately used by GCN, the edges form the paths over which information from neighboring nodes are aggregated by GCN before being used for the final label prediction. The explained heuristic tries to directly estimate the usefulness of such an aggregation for the target prediction task, for each pair of nodes. If the aggregation seems to be useful then an edge is added with an appropriate weight, while the edge is dropped otherwise. 

\begin{figure}[H]
\includegraphics[width=1\linewidth]{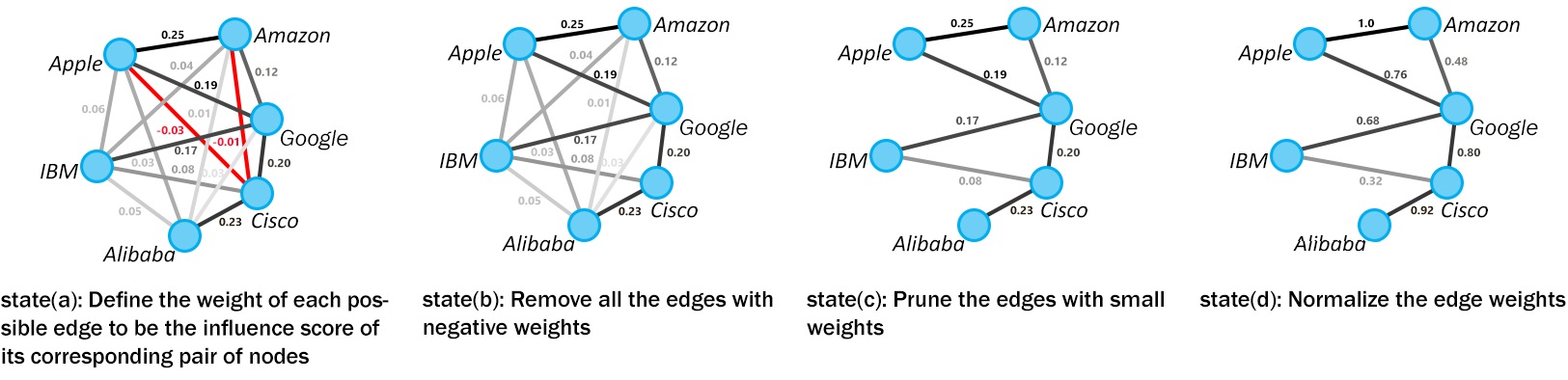}
\caption{Graph construction steps}
\label{graph_con}
\end{figure}

In the final step, GCNET removes the edges with the lowest weight values one by one, and continues to do so until removing the next edge will make the graph disconnected. This sparsification process keeps the strong connections in the graph and removes the unreliable low-weight edges that may introduce noise to the later steps of the algorithm. In addition, using a sparser graph improves the efficiency of the GCNET in learning the final label prediction model. The steps of the influence graph construction procedure is summarized in Algorithm \textcolor{mycolor}{2} and Fig. \ref{graph_con}.

%\subsubsection{Network preparation}
The explained influence network structure can be used to represent each single day of the market. Basically the embedding and prediction steps of the GCNET can also be applied on the network model of the target day, we decided to use a set of five networks to represent each day. In this approach we construct five structurally similar networks, one for the target day and four others for the four days before that. The adjacency matrix of all these networks are similar and the same as the constructed influence network, each stock has a corresponding node in each of those graphs with a different attribute vector belonging to a different day (Figure \textcolor{mycolor}{2}). Using the network representations of the recent days, in addition to the target day, enriches the input data of the embedding and prediction step and allows it to consider the recent past states and movements of the stocks as well to generate more general embeddings and hopefully a more stable prediction model.

\begin{figure}[H]
\centering
\includegraphics[width=.9\linewidth]{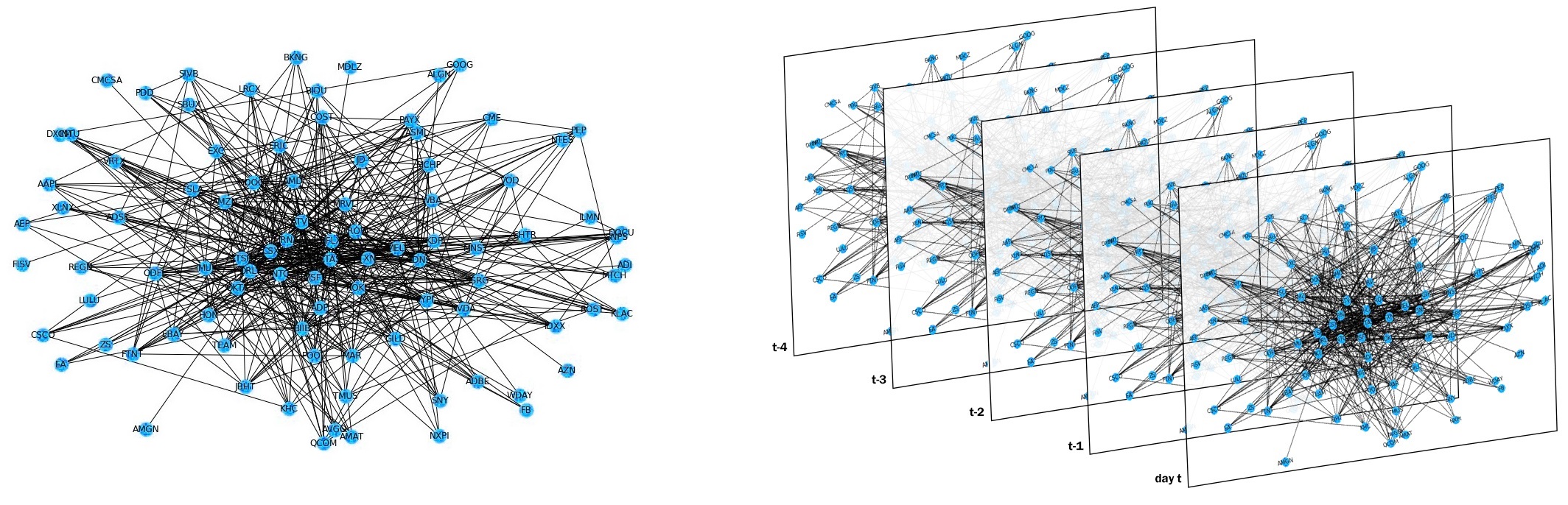}
\caption{ A sample influence graph generated from the dataset used in this paper to represent the stock relations (left). This structure is used by GCNET to form a graph representation of the five recent days (right) that undergoes a separate labeling process for each target day t.} \label{graph_str}
\end{figure}

\begin{algorithm}[H]
\hspace*{\algorithmicindent} \textbf{Input :}\\\hspace*{\algorithmicindent}\hspace*{\algorithmicindent}\textbf{Dataset} (Price history for m stocks) \\
\hspace*{\algorithmicindent} \textbf{Output :}\\\hspace*{\algorithmicindent}\hspace*{\algorithmicindent}\textbf{Graph G(V,E)}  
\begin{algorithmic}
\vspace{-3mm}
\\\hrulefill
\State $V\gets \lbrace \rbrace$, $E\gets \lbrace \rbrace$
\For{each stock $s_i$}
\State $V\gets V\bigcup \lbrace n_i\rbrace$ 
\State $X^{T}_{s_{i}}, X^{V}_{s_{i}} \gets split \hspace{0.1cm} data$(matrix of features of $s_i$) \Comment{Set Train and Validation part of data}
\State $F^{T}_{s_{i}}, F^{V}_{s_{i}} \gets split \hspace{0.1cm} data$(vector of past price fluctuations of $s_i$)
\EndFor
\For{each pair of stocks $(s_i,s_j)$}

\State $Acc_i \gets QDA_{i}.fit(X^{T}_{s_{i}}, F^{T}_{s_{i}}).score(X^{V}_{s_{i}}, F^{V}_{s_{i}})$
\State $Acc_j \gets QDA_{j}.fit(X^{T}_{s_{j}}, F^{T}_{s_{j}}).score(X^{V}_{s_{j}}, F^{V}_{s_{j}})$
\State $X_{Processed} \gets 1/2(X_{s_i} + X_{s_j})$\Comment{Averaging the two feature vectors to combine information}
\State $X^{T}_{Processed}, X^{V}_{Processed} \gets split\hspace{0.1cm}data(X_{Processed})$
\State $Acc_{ij} \gets QDA_{P_{ij}}.fit(X^{T}_{Processed}, F^{T}_{s_{i}}).score(X^{V}_{Processed}, F^{V}_{s_{i}})$
\State $Acc_{ji}  \gets QDA_{P_{ji}}.fit(X^{T}_{Processed}, F^{T}_{s_{j}}).score(X^{V}_{Processed}, F^{V}_{s_{j}})$
\State $Influence_{i,j} = 1/2((Acc_{ij} - Acc_i)+(Acc_{ji} - Acc_j))$ \Comment{Average prediction improvement for the two stocks}
\If{$Influence_{i,j} > 0$} 
\State $E\gets E\bigcup \lbrace e_{i,j}\rbrace$
\State $w_{i,j}\gets Influence_{i,j}$ \Comment{Set the weight of $e_{i,j}$}
\EndIf
\EndFor
\\ $E \gets sort_{asc}(E)$ \Comment{Sort the edges in ascending order of weights}
\For{each edge $(e_{i,j})$ in $E$} \Comment {Edges with smaller weights are processed first}
\If{$G$ is connected}
\State $E\gets E - \lbrace e_{i,j}\rbrace$
\State $e_{last} \gets e_{i,j}$
\Else
\State $E\gets E \bigcup \lbrace e_{last}\rbrace$ \Comment{Add edge that makes graph connected}
\State \textbf{return} $(G(V,E))$
\EndIf
\EndFor
\caption{GenerateInfluenceGraph}
\end{algorithmic}
\label{alg:InfluGConst}
\end{algorithm}

\subsection{Label assignment process}
As mentioned earlier, GCNET uses a semi-supervised style of learning to generate a label prediction model, that uses a partially labeled graphical representation of data as input. In this setting, the label of a graph node is supposed to represent the next price movement for its corresponding stock, while the label prediction's task is to predict the next price movement for stocks whose labels are unknown. The graph structure that represents each training instance in GCNET, as we explained in the last section, is composed of five structurally identical networks modeling five different days of the market. For the days before the target test day, the next day's price of each stock is already known and is used to label the nodes in those four graphs. However, for the last network, that models the last day of the market, the stocks' next price movements are unknown and need to be predicted by the GCENT. However, before passing the set of five networks to the next step, GCNET assigns initial labels to some of the nodes of this test-day network as well, by making reliable predictions about the future movements of their corresponding stocks using a so-called PLD mechanism. Although those initial labels can be refined and changed by the algorithm in the next steps, assigning those initial labels injects some intuition about the possible future of the market into the graph to improve the performance of the subsequent embedding and prediction steps of GCNET.

\subsubsection{Plausible Label Discovery(PLD)}
\label{PLD}
Here we introduce the PLD algorithm that  assigns some initial plausible predicted labels to a subset of the nodes in the test-day graph.
PLD trains a set $PA$ of agile basic history-based algorithms to predict the labels for the nodes of the graph on a daily basis. 

Suppose that for the prediction of the rise/fall in stock prices for a target day $t$ we have a set of labeled instances for each stock $s_i$ for a period of $T$ days before $t$ in historical data. PLD splits the dataset into training and validation sets, such that the validation set $VS_{s_i}$ contains samples belonging to a period of $d$ days before t and the training set $TS_{s_i}$contains older samples.
For each stock $s_i$, a prediction model is generated by each of the basic algorithms using the training set $TS_{s_i}$.

PLD then scores the predictors of each stock by evaluating their prediction performance over the validation set of that stock. The reason for selecting the most recent labeled instances as the validation set is to make the scores assigned to the predictors more reliable by using the instances that are closer to the target day for evaluation. Also, to put more emphasis on predictions closer to the target day $t$ when evaluating the predictors over the validation set, we use a weighted scoring method that gives higher weights to the accuracy of predictions made for the more recent days. The scoring method has been defined in equation \eqref{eq:SCORE}:

\begin{align}
\label{eq:SCORE}
		\begin{split}
			& score_{s_i}^j = \sum_{i=0}^{d-1} {a_{t-1-i}\times (1-c)^{i}}\\
			where\\
			& a_{l}=
				\begin{cases}
					1 & \text{if predictor j predicts the class of instance $ins_{s_i}^l$ correctly} \\
					0 & \text{otherwise }
				\end{cases} \\
			and\\
			& 0<c\ll 1
		\end{split}
\end{align}

Once all the predictors are scored, we select the predictor with the highest score, $best_{s_i}$, for each stock $s_i$.  GCNET assumes that $best_{s_i}$ will have the most reliable prediction, among predictors of stock $s_i$, for the price movement of $s_i$ on the target day $t$. Here we call this score the predictability score of the stock node $s_i$
(\ref{eq:node_pedictability}):
\begin{align}
\label{eq:node_pedictability}
predictability_{s_i} = score_{s_i}^{best_{s_i}} 
\end{align}

To select a node for initial labeling, in addition to its predictability score, GCNET also considers how dense the graph is in the neighborhood of that node. Intuitively, nodes that are located in denser neighborhoods are possibly better candidates for initial labeling as their label information can be shared through several paths with many other nodes in the later embedding steps of the algorithm.

%In the last step, we want to select nodes for labeling that have as accurately predicted labels as possible and also have the greatest impact on the GCN training process. GCN is basically a semi-supervised model that uses the underlying graph structure and feature vectors of test samples and one label in the GCN process belongs to an embedding of stock and its neighbors.

To evaluate the neighborhood density of each node, we use the weighted local clustering coefficient \citep{saramaki2007generalizations} to define the density score for a node $s_i$ as:

\begin{align}
\label{eq:clustering}
density_{s_i} = C_i = \frac{1}{deg(i)(deg(i)-1)} \sum_{j,k}(\hat{w}_{i,j} \hat{w}_{i,k} \hat{w}_{j,k})^{1/3}
\end{align}
Where the edge weights are normalized by the maximum weight in the network, $\hat{w}_{i,j}=w_{i,j}/max(w)$, and the contribution of each triangle depends on all of its edge weights. This measure assigns high scores to the nodes located at the center of dense neighborhoods and lower scores to the nodes in whose neighborhood strong edges are rare.  
The final score assigned to each node is the defined by the product of its predictability and density score as in (\ref{eq:node_privilege}):

\begin{align}
\label{eq:node_privilege}
privilege_{s_i} = predictability_{s_i}*density_{s_i}
\end{align}

In the final step, n\% of the nodes with the highest node privilege are selected to be assigned initial labels. The assigned label to each selected node is the prediction of its corresponding stock's best predictor,$best_{s_i}$ for the target day: if the stock node $s_i$ is among the top n\% nodes with the highest privilege values and its best predictor predicts a rise/fall in its price it will be labeled with +1/-1. However, if the node's privilege score is not among the top n\% scores then it will be left unlabeled. 
The main steps of the PLD procedure are presented in Algorithm \textcolor{mycolor}{4}.

\begin{algorithm}[H]
\hspace*{\algorithmicindent} \textbf{Input :}\\\hspace*{\algorithmicindent}\hspace*{\algorithmicindent}\textbf{Training Data} (training instances for all stocks) \\
\hspace*{\algorithmicindent}\hspace*{\algorithmicindent}\textbf{Graph G(V,E)}\\
\hspace*{\algorithmicindent} \textbf{Output :}\\\hspace*{\algorithmicindent}\hspace*{\algorithmicindent} \textbf{Partially labeled graph G(V,E)}
\begin{algorithmic}
\vspace{-3mm}
\\\hrulefill
\State $PA \gets\lbrace pa^1, pa^2,...,pa^k\rbrace$\Comment{Prediction algorithms}
\For{each stock $s_i$}
 \State $VS_{s_i}\gets\lbrace ins_{s_i}^{t-1}, ins_{s_i}^{t-2}, ...,ins_{s_i}^{t-d}\rbrace$ \Comment{d instances for validation}
 
\State $TS_{s_i}\gets\lbrace ins_{s_i}^{t-d-1}, ins_{s_i}^{t-d-2}, ...,ins_{s_i}^{t-T}\rbrace$ \Comment{The rest for train}

 \For{each algorithm $pa^j$ in $PA$}
   \State $score_{s_i}^j\gets pa^j.train(TS_{s_i}).evaluate(VS_{s_i})$
   \Comment{use $pa^{j}$ to train and evaluate a predictor for stock $s_i$ }
 \EndFor
 
% \State $score_{s_i}^{best}\gets \max_k score_{pr_{s_i}^k}$
 
 \State $best_{s_i}\gets$ The best trained predictor for $s_i$
 \Comment{The one with the highest score}
 \State $predictability_{s_i}\gets  score_{s_i}^{best{s_i}}$

 % %PA(\argmax_k score_{pa_{s_i}^k})$\Comment{Best predictor for stock $s_i$}
%\State $BP \gets BP \bigcup \lbrace best_{s_i}\rbrace$ 
 
\State $density_{s_i}\gets weighted \_local \_ clustering \_ coefficient(G, node_{s_i})$
 
\State $privilege_{s_i} \gets predictability_{s_i} * density_{s_i}$
 %NP \bigcup \lbrace score_{s_i}^{best{s_i}} * C_{s_i}$ %\Comment{Node privilege for stock $s_i$}

 \EndFor
 
 \State $TP\gets$ Top $n\%$ stocks $s_i$ with the highest $privilege_{s_i}$\Comment{The selected nodes to be assigned initial labels}
 %members of $NP$
 
\For {each  stock $s_k$ in $TP$}
   \State $lablel_{s_k}\gets$ price fluctuation predicted by predictor $best_{s_k}$ for the target day t of stock $s_k$
\EndFor
\State return (G(V,E))\Comment{Partially labeled Graph}
\caption{PLD procedure for the target day \textbf{t}}
\end{algorithmic}
\label{alg:PLD}
\end{algorithm}

The set of the agile basic algorithms we have used in PLD includes neural network, Linear Discriminant Analysis, Decision Tree, Naive Bayes, and Quadratic Discriminant Analysis, Random forest. The input features used to represent the instances for training these models are summarized in Table \ref{tab:input}.

\setlength{\tabcolsep}{20pt}
\begin{table}[H]
\centering
\caption{Technical indicators and price information used by basic prediction algorithms in PLD}
\label{tab:input}
\small
\begin{tabular}{|l|l|l|l|}
\hline
Indicator & Description & Indicator & Description
\rule{0pt}{4mm}%
\\[1.5mm]
\hline
\hline
OP & Open price  &  RSI-S  & RSI indicator signal  
\\
HP & High price &  BB-S  & Bollinger bands indicator signal 
\\
LP & Low price &  MACD-S  & MACD indicator signal 
\\
CP & Close price &  SAR-S  & SAR indicator signal 
\\
Volume & Trading volume &  ADX-S  & ADX indicator signal 
\\
CCI & Commodity Channel Index &  S-S  & Stochastic indicator signal 
\\
SAR & Stop and reverse index &  MFI-S  & MFI indicator signal 
\\
ADX & Average directional movement  &  CCI-S  & CCI indicator signal 
\\
MFI & Money flow index &  V-S  &Sign(Volume -Avg(last 5 days)) 
\\ 
RSI & Relative strength Index &  CPOP-S  & Sign(CP-OP) 
\\
SK & Slow stochastic \%K &  CPCPY-S  &  Sign(CP-Closing price yesterday) 
\\
SD & Slow stochastic \%D & &
\\
\hline
\end{tabular}
\end{table}

Please note that the explained model training and evaluation process is repeated for each target day, so, the nodes that are labeled by PLD, are not necessarily the same for different target days.

\subsection{Price Movement Prediction }
In this phase, node features are added to the partially labeled graph of stocks generated in previous steps and the resulted graph is used to train a Graph Convolutional Network model for prediction of the stock price movements as final node labels of the graph. We first present a brief background about the Graph Convolutional Network model and then we will explain the details of model training procedure.

\subsubsection{GCN Background} 
Graph Convolutional Network (GCN) is a framework for representation learning in graphs and is derived from graph signal processing theory. GCN can be applied directly on graph structured data to extract informative representations for each node by aggregating information from its neighbors in depth $d$. GCN can be considered as a Laplacian smoothing operator for node features over graph structures. The architecture of GCN consists of a series of convolutional layers. Each layer of GCN integrates information from the neighbors of each node to update its representation according to the structure of all nodes and edges in the graph by the Laplacian matrix.

In spectral theory, the convolution operation is defined in the Fourier domain by computing the eigendecomposition of the graph Laplacian. The operation can be defined as the multiplication of a signal $ x \in \mathbb{R}^{N}$ (a scalar for each node) with a filter $ \textsl{g}_\theta =\text{diag}(\theta)$ parameterized by $ \theta \in \mathbb{R}^{N}$:

\begin{equation}
\label{eq:con_1}
	  \textsl{g}_\theta \star x = \textbf{U} \textsl{g}_\theta (\Lambda) \textbf{U}^T x
\end{equation}

where $\textbf{U}$ is the matrix of eigenvectors of the normalized graph Laplacian (L), with a diagonal matrix of its eigenvalues $\Lambda$. Equation (\ref{eq:con_1}) describes a spectral convolution filter $g_\theta$ used for graph data $x$. This operation results in potentially intense computations. 
\cite{hammond2011wavelets} approximated spectral filters in terms of Chebyshev polynomials and \cite {kipf2016semi} limit the layer-wise convolution operation to K = 1 to alleviate the problem of overfitting on local neighborhood structures for graphs with very wide node degree distributions and Introduce the Graph Convolution Network.
Also, GCN simply transform Equation (\ref{eq:con_1}) as a fully connected layer with a built-in convolution filter, which is written as the following equation:

\begin{align}
\label{eq:GCN}
H^{l+1} = ReLU(\widehat{A}H^{l}W^{l})
\end{align}

where $\widehat{A} = \tilde{D}^{-\frac{1}{2}} \tilde{A} \tilde{D}^{-\frac{1}{2}}$. Here, $\tilde{A} = A +I_{N}$, in which $\tilde{A}$ is the adjacency matrix of the graph $G$,  with added self-connections, IN is the identity matrix, $\tilde{D} = \sum_{j} \tilde{A}_{ij} $ and $W^{l}$ is a layer-specific train able weight matrix.

\subsubsection{Training and predicting Procedure}

%-------------------------------------------------
Each node of the graphs is represented by an initial feature vector that includes a set of technical indicator signals and the stock's last day price information as features. The complete set of features used for representing each node has been summarized in Table \ref{tab:nodeinput}.
To train the label prediction model on this graph, we use a GCN with three convolution layers. 
%The input to the GCN model contains two elements: initial feature vector X and graph's adjacency matrix A. 
The first hidden layer $H^{0}$ receives the original node features (feature matrix X). Our model runs 3 iterations of updates according to Equation \eqref{eq:GCN} to generate the final output node embeddings and the overall process of generating the output can be defined as:

\begin{align}
\label{eq:GCN_model}
Y = softmax(\widehat{A} ReLU(\widehat{A}ReLU(\widehat{A}XW^{0})W^{1})W^{2})
\end{align}

where $W^{0} \in \mathbb{R}^{|X|\times C_{0}}$ is an input-to-hidden weight matrix for a hidden layer with $C_{0}$ feature maps and $W^{1} \in \mathbb{R}^{C_{0}\times C_{1}}$ is a hidden-to-hidden weight matrix with $C_{1}$ feature maps and $W^{2} \in \mathbb{R}^{C_{1}\times C_{2}}$ is a hidden-to-output weight matrix. The value of $C_{2}$ for prediction of the two classes (rise and fall) is 2. We train the network using cross-entropy loss over predictions for all stocks modeled in the network:
\begin{align}
\label{eq:LOSS}
Loss = - \sum_{i \in Z_{L}} \sum^{F}_{f=1} Z_{if} lnY_{if}
\end{align}

where $Z_{L}$ is the set of stocks with plausible labels, $F$ is the class of stock price movement direction.

Considering the number of features and training examples, we used 4 channels for the first and second layers and 2 channels to predict 2 classes in the third layer.
The activation function for the first and second layers is ReLU and for the third layer is Sigmoid.

\setlength{\tabcolsep}{20pt}
\begin{table}[H]
\centering
\caption{Features used to represent nodes in the graph}
\label{tab:nodeinput}
\small
\begin{tabular}{|l|l|l|l|}
\hline
Indicator & Description & Indicator & Description
\rule{0pt}{4mm}%
\\[1.5mm]
\hline
\hline

RSI-S  & RSI indicator signal  & MFI-S  & MFI indicator signal
\\
BB-S  & Bollinger bands indicator signal & CCI-S  & CCI indicator signal
\\
MACD-S  & MACD indicator signal & V-S  &Sign(Volume -Avg(last 5 days)) 
\\
SAR-S  & SAR indicator signal & CPOP-S  & Sign(CP-OP) 
\\
ADX-S  & ADX indicator signal & CPCPY-S  &  Sign(CP-Closing price yesterday)
\\
S-S  & Stochastic indicator signal & &
\\
\hline
\end{tabular}
\end{table}

The GCN model is trained separately for each test day $t$ during the test period. As we explained before, five graphs representing days $t-4$ to $t$ are used for training the prediction model for the target day $t$, among which graphs for days $t-4$ to $t-1$ are fully labeled and the graph representing the target day $t$ is partially labeled. Each of these graphs is used in turn, in chronological order, to train the model for $n$ epochs, i.e. GCNET first trains the GCN model using the graph of the day $t-4$ for $n$ epochs, then it continues the training process for another $n$ epochs using the graph of the day $t-3$ and so on (the value of $n$ is set using the validation data). So, GCNET tries to train an initial prediction model using the fully-labeled graphs of the previous days. The labels of those graphs are reliable as they represent the actual price fluctuations observed in the corresponding days, and by using them GCNET is able to train a stable initial model that is consistent with the behavior of the market in past few days. That initial model is then adjusted in the last $n$ training epochs using the partially labeled graph of the target day $t$ that represents the most recent state of the market as well as some initial predictions about its future. By processing the graphs in chronological order GCNET practically puts an emphasis on the more recent information, as they are used in the late steps of training in which the GCN model is fine-tuned and finalized.

Using the explained process, GCNET trains a GCN model to predict the labels for all the nodes of the day $t$ graph, whether they have initial labels or not. The prediction of the GCNET about the next price movements of each stock in the market can then be inferred from the final label predicted for its corresponding node in the day $t$ graph. GCNET repeats these model training and label prediction steps for each prediction day in the data. Figure \textcolor{mycolor}{3} shows the component architecture and steps of GCNET.

\begin{figure*}
\includegraphics[width=1\linewidth]{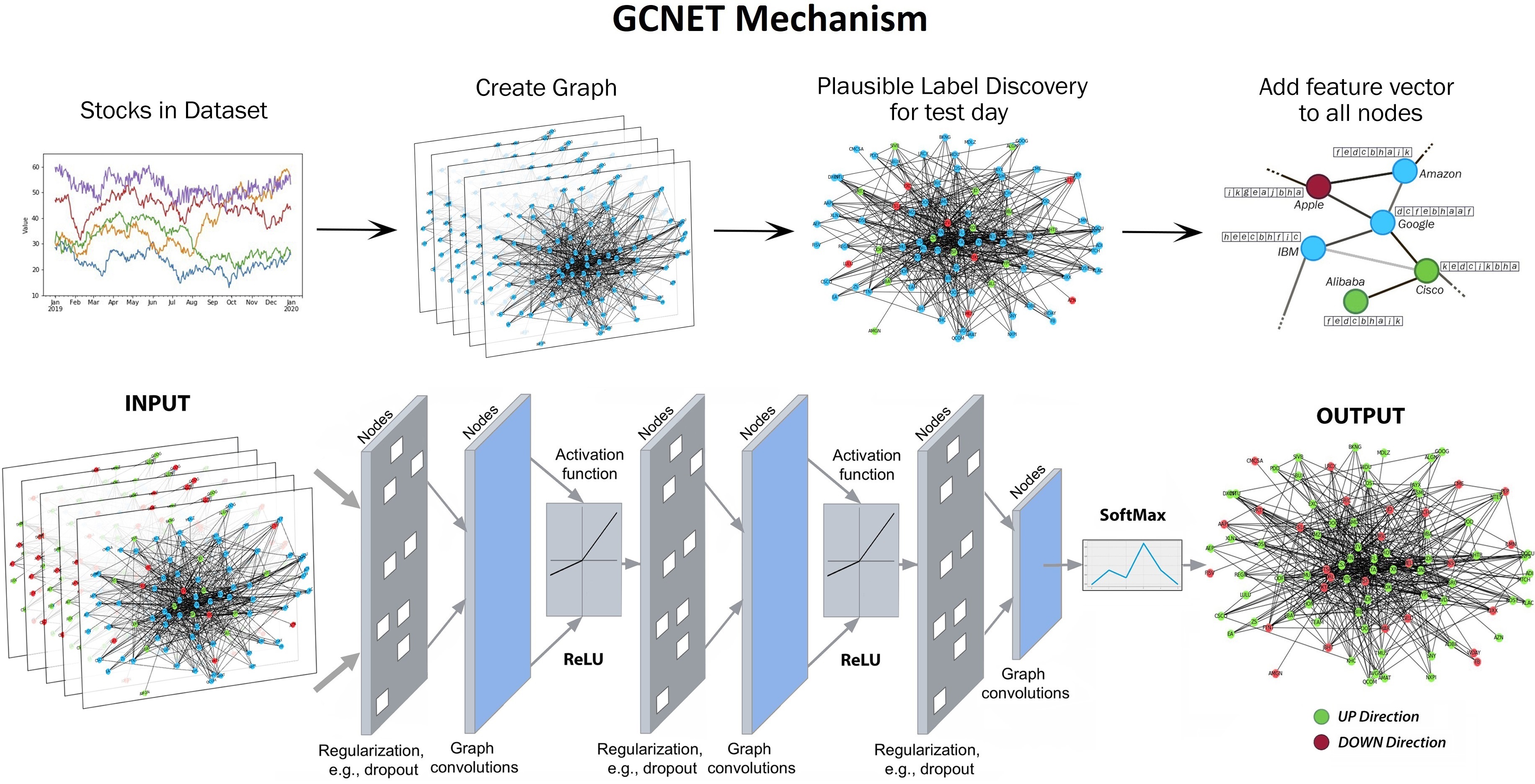}
\caption{Architecture of the components of the GCNET model}
\label{scheme}
\end{figure*}

\subsection{Complexity analysis}
\label{complexity}

GCNET algorithm is comprised of three phases: creating the network, running PLD, and training GCN. In the following, we use $m$ and $n$ to denote the number of stocks/nodes and the number of training data samples for each stock. 

To create the influence network, first, the influence value between each pair of stocks is calculated by training the required QDA prediction models as we explained in section \ref{infnet}. QDA is a variation of LDA algorithms with the time complexity of $O(knf+ k^3)$ where $f$ is the number of features, and $k = \min (f,n)$. When the number of training samples and the number of features is small, as they are in this work, this algorithm runs fast. In our implementations, we use a small set of features and $f < n$, so the time complexity is $O(nf^2+ f^3)$. The number of models that must be trained to construct the entire graph is $O(m^2)$ and the total computation complexity for generating the complete influence network is $O(m^2(nf^2+ f^3))$. The computational cost of sparsifying each graph is $O(m^2)$.

The computational time complexity of the PLD phase depends on the set of basic algorithms we select to use.  We have applied simple algorithms among which random forest has the highest time complexity that dominates the running time of this module. The Time complexity for building a complete decision tree is $O(f \hspace{0.5mm} n \hspace{0.6mm} log \hspace{0.5mm} n)$, where $f$ is the number of features used by the model. So, the computational complexity of the random forest algorithm is $O(k \hspace{0.5mm} f \hspace{0.5mm} n \hspace{0.6mm} log \hspace{0.4mm} n)$, where $k$ is the number of trees generated by the model \citep{hassine2019important}. Since the PLD algorithm trains its models for each stock separately, its model training time complexity for the configuration used in this study would be $O(k \hspace{0.5mm} f \hspace{0.5mm}m\hspace{0.5mm} n \hspace{0.6mm} log \hspace{0.4mm} n)$. 
One can decide to train the basic algorithms for every prediction day, or instead, use the trained models for a few days before retraining or tuning them. In this study, we use a set of simple algorithms to keep the computational time low, and retrain them for each prediction day.

The time complexity of GCN is $O(L \hspace{0.4mm} ||\widehat{A}||_0 \hspace{0.4mm} f + L \hspace{0.4mm} N f^2)$, where $||\widehat{A}||_0$ is the number of nonzero values of the adjacency matrix and $f'$, $L$ and $n$ denote the size of feature vectors used by GCN, the number of GCN layers  and the number of the nodes(i.e. stocks) respectively \citep{wu2020comprehensive}. The GCN training step is repeated for each prediction day while the trained model is used to predict the labels for all stock nodes in the graph.

Regarding the fact that the underlying graph used by GCNET is rather small and sparse, the running time for its periodical construction and daily analysis is low. So in practice, the training time of GCNET is almost linearly proportional to the running time of the most time-consuming algorithm used in PLD. If, as we did in this study, simple and efficient algorithms are used in PLD, the GCNET algorithm will be more time-efficient than most of the state-of-the-art deep models used by single stock prediction systems.

\section{Experiments}
\label{experiment}

In this section, we first explain the data and the pre-processing step, and then we present the evaluation process and the results of the GCNET model.

\subsection{Data description}
\label{data}

To evaluate the performance of GCNET and the baseline algorithms we used a large dataset of the famous stocks in the Nasdaq market. 
Our dataset contains the historical data for 100 stocks with the largest market capitalization in the Nasdaq index from 01/01/2011 to 01/01/2021. The price data has been obtained from Yahoo Finance. Each daily record of data for each stock contains open, low, high, close, adjusted close, and volume values. The historical data before each test day is used to train the model. The PLD models and the semi-supervised label prediction mechanisms are retrained for every prediction day. The data in the 09/15/2020 to 09/29/2020 interval is used for validation and to set the hyper-parameters. The test period is from 09/30/2020 to 12/30/2020 which includes 5952 records. 

%Our dataset is available at the following %address\footnote{\href{https://github.com/ut-kdd/GCNET-Dataset}{https://github.com/ut-kdd/GCNET-Dataset}}.

\subsection{Implementation and Parameters of GCNET}
To keep the graph structure up to date, the graph can be rebuilt in t-day intervals. If the training and validation dataset using which the edge weights are calculated are large enough, the resulted graph is expected to perform well for a long period of time, however, in our experiments, we reproduce the graph every 30 days. Fig. \ref{graph_str} shows a sample graph that is generated for the dataset used in this paper.
We apply Adam optimizer \citep{da2014method} to train the GCN and to avoid over-fitting we applied L2 regularization and dropout techniques. The weights of the GCN have been initialized by Glorot uniform initializer \citep{glorot2010understanding}. For GCN implementations we used Keras \citep{chollet2015keras} and Spektral \citep{grattarola2020graph} libraries.
The number of top \%n nodes that are labeled by the basic models is set using the validation data. Other parameters like learning rate and dropout rate have been set to widely-used values in the literature. %The source code for our proposed model is publicly available\footnote{\href{https://github.com/ut-kdd/GCNET-Code}{https://github.com/ut-kdd/GCNET-Code}}.

\subsection{Evaluation Metrics}
Following previous work in stock prediction domain \citep{WU2022405,kia2018hybrid}, we adopt the standard measure of accuracy and Matthews Correlation Coefficient (MCC) as evaluation metrics Given the confusion matrix
$\begin{pmatrix}tp & fn\\ fp & tn\end{pmatrix}$
containing the number of samples classified as true positive, false positive, true negative and false negative, Accuracy and MCC are defined as:

\begin{equation}
ACC = \frac{tp + tn}{tp + tn + fp + fn}
\end{equation}

\begin{equation}
MCC = \frac{tp \times tn - fp \times fn}{\sqrt{(tp + fp)(tp + fn)(tn + fp)(tn + fn)}}
\end{equation}

\subsection{Baselines}
We compare our model with the following baselines, which include network-based models, deep learning models, artificial neural network models, and models that use feature selection:
\begin{itemize} 
\item \textbf{ALSTM:} This is a deep model, which uses a dual-stage attention-based recurrent neural network. In the first stage, the attention mechanism is applied to input features and is used to extract the relevant representation at each time step by referring to the previous hidden state. In the second stage, a temporal attention mechanism is applied to select relevant encoder hidden states across all time steps. Finally, an LSTM neural network is used to predict stock movement. \citep{qin2017dual}.

\item \textbf{STOCKNET:} A variational auto-encoder that combines price and text information. Price features are modeled sequentially and The text is encrypted using a hierarchical attention mechanism in a multi-day sequence \citep{xu2018stock}.

\item \textbf{HATS:} A hierarchical graph attention method that uses graph modeling to weigh different relationships between stocks. they use a relational modeling module with initialized node representations. Their model selectively aggregates information on different relation types and adds the information to the representations of each company. they use only historical price data. \citep{kim2019hats}.

\item \textbf{CNN-pred:} A deep CNN model that uses a diverse set of financial variables, including technical indicators, stock indices, futures contracts, etc. Their model constructs three and two-dimensional input tensors. Then, the input tensors are fed into a specified CNN model to make predictions. They use a fixed combination of the target market's time series as well as a few related time series as input to the model to predict the future of the target time series. In our experiments, we used data from three closest neighbors of the target stock in the correlation graph as the related time series whose data is fed to the model\citep{hoseinzade2019cnnpred}.

\item \textbf{Adv-LSTM:} 
In this work, the authors introduced a deep model using an Adversarial Attentive LSTM mechanism, which leverages adversarial training to simulate the stochasticity during model training. They advised employing adversarial training to improve the generalization of a neural network prediction model because the input features to stock prediction are typically based on stock price, which is essentially a stochastic variable and continuously changed with time by nature. Adversarial learning trains a classification model to defense adversarial examples, which are intentionally generated to perturb the model. They added perturbations to their stock data and train the model to work well under small yet intentional perturbations. In our experiment, we have tried to tune the hyperparameters, the number of hidden units, lag size, and weight of the regularization term \citep{feng2019enhancing}.

\item \textbf{exLift+DiMexRank+(PLD):} exLift + DiMexRank is a network-based model for prediction of market indices that encrypts relationships between markets' indices using Association rule learning and predicts the direction of market index movement using PageRank algorithm. The initial labeling step of the original algorithm is designed for a specific market prediction task and cannot be applied in the domain of this study. Therefore, to use it as a baseline algorithm, we apply our suggested PLD method for the initial labeling step in exLift+DimexRank, and keep the rest of the model intact.  \citep{kia2020network}.

\item \textbf{Price graphs:} a graph-based model that transforms time series into visibility graphs, then, structural information and referring to temporal point associations are extracted from the mapped graphs by a graph embedding method. Price graphs use the attention-based layers and a fully connected layer for stock trend prediction \citep{WU2022405}.

\end{itemize}

\subsection{Results and discussion}

In this section, we compare our model to the baseline models mentioned in the previous section. 
Our evaluations were performed on a large set of stocks in the Nasdaq market, and Tables  \ref{tab:RESULT_1} summarizes the average performance of GCNET and baseline algorithms on 10 independent experiments. 
%We have evaluated two different versions of GCNET, GCNET-inf uses the influence network as its underlying graphical model while GCNET-cor uses the correlation network. Also, to demonstrate the importance of our labeling process, we evaluated a GCN model without the PLD method.  

As it can be seen, GCNET outperforms all baselines, while HATS, Price graphs, and exLift+ DimexRank+(PLD) achieve the highest prediction performances among baselines. The significant superiority of GCNET over baseline algorithms most of which use different sources of information and diverse sets of features as well as sophisticated prediction techniques demonstrates the power of the proposed graph-based prediction algorithm.
As another important observation, graph-based models, especially GCNET, achieve significantly higher values of MCC compared to other algorithms. MCC, or Matthew's Correlation Coefficient, is designed to summarize the confusion matrix. A high value in the MCC measure reflects the power of a model in successfully predicting both ascent and descent classes. 
One reason for this superiority can be the fact that graph-based techniques exploit the graph structure to infer a consistent set of predicted final labels. In such a setting, if an initial prediction is mistakenly biased towards one of the classes for a certain stock, it can be corrected by the subsequent network-based information-aggregation steps.

The satisfactory performance of exLi-ft+DimexRank+(PLD), as a simple graph-based algorithm, affirms the usefulness of the graphical modeling in this domain, even if a simple label prediction mechanism is applied in that algorithm. Also, the superiority of our models over exLift+DiMexRank+(PLD) (that uses random walk analysis to assign labels to the nodes), shows the positive effect of formulating the price movement prediction problem as a model-based label prediction task. As we mentioned before, discovering and combining latent information from different stocks, whose relations are modeled as a graph structure, is the core ability of GCN that we exploited in our suggested algorithm, and as the experimental observations show, it can lead to significant improvement of the prediction performance.

Also, Price graphs, which is a powerful and recently introduced graph-based deep model, has achieved an accuracy of 55.21, which is significantly lower than the performance of GCNET. The Price graphs, unlike GCNET, uses the visibility graphs to model the relations between samples of a single stock over time, and the superiority of GCNET shows that our introduced approach for graph-based analysis of different stocks relations seems to be a more reliable and effective technique in the prediction of stock price fluctuations.

\setlength{\extrarowheight}{4pt}
\setlength{\tabcolsep}{5pt}
\begin{table}[H]
\centering
\caption{Comparison of GCNET model result with baselines}
\label{tab:RESULT_1}
\begin{tabular}{lcc }
\hline
Model & Accuracy & MCC
\rule{0pt}{4mm}%
\\[1.5mm]
\hline
\hline

ALSTM \citep{qin2017dual}  & 52.63 $\pm 0.04 $  & -0.002 $\pm 0.001 $ 
\\ 
STOCKNET \citep{xu2018stock} & 54.74 $\pm 0.27 $  & 0.0375 $\pm 0.009 $ 
\\

HATS \citep{kim2019hats} & 55.12 $\pm 0.21 $ & 0.0618 $\pm 0.007 $ 
\\
CNN-pred \citep{hoseinzade2019cnnpred}  & 53.42 $\pm 0.62 $  & 0.0366 $\pm 0.015 $ 
\\ 
Adv-LSTM \citep{feng2019enhancing}  & 54.66 $\pm 0.45 $  & 0.0412 $\pm 0.011 $ 
\\ 
exLift+ DiMexRank+(PLD)\citep{kia2020network}  & 54.03 $\pm 0.68 $  & 0.0744 $\pm 0.013 $ 
\\ 
Price graphs \citep{WU2022405}   & 55.21 $\pm 0.32 $  & 0.0821 $\pm 0.008 $ 
\\ 
\hline
\textbf{GCNET} (our model) & \textbf{56.71} $\pm 0.49 $  & \textbf{0.1013} $\pm 0.011 $
\\
\hline
\end{tabular}
\end{table}

\subsubsection{Model components analysis}
In this subsection, we are going to discuss the importance of model components separately and provide more experiments for each component. First, we investigate the role of the influence network and then we discuss the node selection mechanism and GCN.

\subparagraph{Network}
The influence network discovers about 2300 edges during the weighting process, and after pruning the edges with low weight, the graph retains the effect of about 800 edges, which achieved an average of 3.67\% increase in forecast accuracy in network validation data. 

To see how GCNET performs when using another approach for modeling the stock relations, we replaced the influence network  with a correlation network, that,
as we explained in the literature review section, is the most commonly used network structure for predicting stocks.
We created a network of stocks in which the edges represent the correlations among stocks past prices. We then used the same pruning technique applied for the influence network so that the edges connecting the strongly correlated stock nodes remain in the network and the pruned network is still connected. We kept the rest of the GCNET algorithm intact and evaluated its performance in prediction of the stock price movements. The results are summarized in the Table \ref{tab:RESULTs_4}:

\setlength{\extrarowheight}{4pt}
\setlength{\tabcolsep}{5pt}
\begin{table}[H]
\centering
\caption{Comparison table of correlation and influence network performance.}
\label{tab:RESULTs_4}
\begin{tabular}{lc}
\hline
method & Accuracy
\rule{0pt}{4mm}%
\\[1.5mm]
\hline
\hline
GCNET with Correlation graph & 55.79
\\
GCNET with Influence graph & 56.71
\\
\hline
\end{tabular}
\end{table}

As the results show, the original version of GCNET outperforms the one using the  well known and widely used correlation graph with a significant improvement. 
Another interesting observation is that even the correlation based version of GCNET is still superior to other baselines as reported in Table \ref{tab:RESULT_1}, that demonstrates the effectiveness of the introduced graph-based framework compared to other state of the art approaches for the prediction of stock price movements.

%The use of visibility graphs is another category in stock models that uses complex network concepts. These models usually consider the relationships between samples of one stock over time in a graph structure, although, their representations do not include the relationship between the stocks and just focus on the chart of stock. The Price graphs is a very powerful deep model in this field that models the relationships between samples over time. However, the GCNET model has performed significantly better.

%\subsubsection{Label assignment analysis}
\subparagraph{Semi-supervised training}
As the main part of our model, GCN predicts the next day by combining neighbors’ information and considering labeled nodes for stock movement.
To demonstrate the importance of using the GCN algorithm in a semi-supervised approach, we compared our model with two different supervised versions in both of which the model  is trained only with the graphs from previous days whose nodes have known labels (This means that the last day's graph whose labels are unknown and in the original setting, was partially labeled using the predictions of the PLD mechanism is no longer available to the GCN model). In the first supervised  model, GCN model trained using the fully-labeled networks of the past days is directly used to predict the target day's price movements while in the second model, we fed the created embeddings by the GCN to an LSTM model for prediction of the target day's price movements. The results are shown in table \ref{tab:RESULTs_5}:

\setlength{\extrarowheight}{4pt}
\setlength{\tabcolsep}{5pt}
\begin{table}[H]
\centering
\caption{Comparison table of GCN in supervised framework and GCNET.}
\label{tab:RESULTs_5}
\begin{tabular}{lc}
\hline
method & Accuracy
\rule{0pt}{4mm}%
\\[1.5mm]
\hline
\hline
Supervised GCN & 55.08
\\
Supervised GCN+LSTM & 55.46
\\
GCNET (Semi-Supervised) (our model) & 56.71
\\
\hline
\end{tabular}
\end{table}

The table \ref{tab:RESULTs_5} shows that the use of GCN in a supervised framework reduces the performance of the model, and confirms that our suggested semi-supervised approach for extracting embeddings leads to a superior label-prediction performance. The reason is that in the suggested semi-supervised approach for embedding, the most recent information about the state of each stock is considered during the GCN weight training and learning process over the last-day's partially labeled graph \citep{kipf2016semi}.

%\subparagraph{Node selection}
\subparagraph{Label assignment}

As we explained before, the algorithm uses a pool of basic agile predictors and by network analysis techniques, selects the most effective nodes for labeling as well as the best available predictions to be used as their label.
To see the effect of using the pool of predictors mechanism for assigning accurate initial labels, we evaluated the performance of the model when using each single member of the pool, instead of the original ensemble technique, for label assignment. The results are summarized in Table \ref{tab:pld} and it can be seen that using the original ensemble labeling technique has led to a significant improvement in the GCNET's overall performance. It can also be seen that among the basic algorithms used by PLD, Random Forest has the highest prediction performance, which is an interesting observation, as PLD and Random Forest are both ensemble methods.

\setlength{\extrarowheight}{4pt}
\setlength{\tabcolsep}{20pt}
\begin{table}[H]
\centering
\caption{Comparison of different techniques for assigning labels to supervised nodes in test day in GCNET}
\label{tab:pld}
\small
\begin{tabular}{lc}
\hline
The method used for initial label assignment &  Overall accuracy of \textbf{GCNET}  
\rule{0pt}{4mm}%
\\[1.5mm]
\hline
\hline
Support Vector Machine  & 52.76 
\\
Fisher Discriminant Analysis  & 53.62
\\
Decision Tree &  53.89 
\\
ANN & 54.03
\\
LSTM & 54.62
\\
Linear Discriminant Analysis & 54.94
\\
Random Forest & 55.11
\\
\hline
Plausible Label Discovery (PLD) & 56.71
\\
\hline
\end{tabular}
\end{table}

As we know, the GCNET assigns initial labels to a subset of the nodes in the graph of the target day (as an initial guess about the future price movements of some stocks) and leaves the rest of the nodes unlabeled. After processing this partially labeled graph, GCNET generates its final predictions for every stock, including both initially labeled and unlabeled ones. To see how the GCNET prediction mechanism improves the initial labels, please see
Table \ref{tab:RESULTs_3}. As it can be seen, on average, only 52.47 of the initial labels have been correct,  while GCNET improves the accuracy of those labels in its final prediction by about 3.5\%. It also achieves even a higher overall prediction accuracy, considering both initially labeled and unlabeled nodes. This is an important observation that shows the performance of GCNET exceeds the prediction performance of PLD, even for the nodes that have initially been labeled by PLD. The reason is that GCNET combines initial predictions with the information reflecting the graph structure and the local node states to refine the initial guesses and make consistent and reliable final predictions. This observation confirms the usefulness of graph construction and analysis techniques used by GCNET.

\setlength{\extrarowheight}{4pt}
\setlength{\tabcolsep}{5pt}
\begin{table}[H]
\centering
\caption{The performance of GCNET in improving the accuracy of its initial predictions}
\label{tab:RESULTs_3}
\begin{tabular}{lcc }
\hline
Data & Accuracy
\rule{0pt}{4mm}%
\\[1.5mm]
\hline
\hline
PLD on initially labeled nodes & 52.47
\\
GCNET on initially labeled nodes & 55.98
\\
GCNET on all nodes & 56.71
\\
\hline
\end{tabular}
\end{table}

Also, GCNET uses a procedure to select some nodes for labeling by considering the density of each part of the graph. To see the effect of node selection procedure, we evaluated the performance of GCNET version in which a random subset of nodes are selected and then the label produced by their best predictors are assigned to them as initial labels. The results are represented in table \ref{tab:RESULTs_6}. It can be seen that our original node selection procedure has a critical role in the overall GCNET performance improvement.

\setlength{\extrarowheight}{4pt}
\setlength{\tabcolsep}{5pt}
\begin{table}[H]
\centering
\caption{The effect of GCNET's node selection procedure on its overall performance}
\label{tab:RESULTs_6}
\begin{tabular}{lcc }
\hline
Data & Accuracy
\rule{0pt}{4mm}%
\\[1.5mm]
\hline
\hline
GCNET \small{with random nodes selected for initial label assignment} & 54.98
\\
GCNET  & 56.71
\\
\hline
\end{tabular}
\end{table}

To report the effect of n, the portion of nodes that are initially labeled by PLD, on the performance of the algorithm, we presented the prediction's performance for different values of this parameter in Figure \textcolor{mycolor}{4}. One advantage of GCN is its outstanding performance when training the model with a rather small set of labeled nodes \citep{kipf2016semi}. However, as shown in this figure, using a very small value for n, reduces the accuracy of the prediction possibly because in such a situation, the information provided by the labeled nodes would not be enough for extracting reliable patterns for prediction. On the other hand, labeling too many nodes by the basic predictors reduces the performance of the algorithm. These observations confirm the positive role of network analysis in boosting the performance of the prediction: Injecting too much information from the history based models into the graph, not only can introduce some noise to the model but also limits the effect of GCN in the label prediction process, that apparently reduces the performance of the algorithm.

\begin{figure}[H]
\centering
\includegraphics[width=0.6\linewidth]{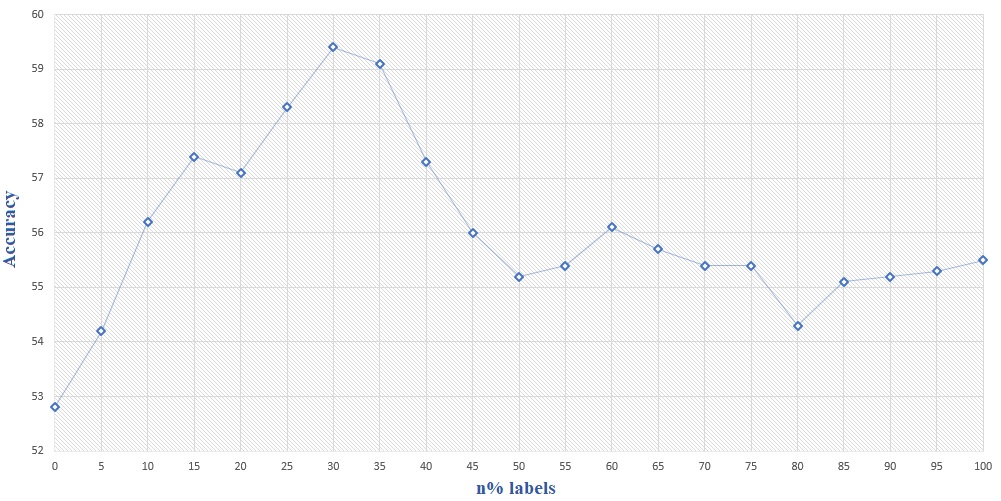}
\label{fig:threshold_valid}
\caption{Prediction accuracy chart based on different label thresholds on validation part of the dataset used in this paper}
\end{figure}

\section{Conclusion}
\label{conclusion}

In this paper, we presented a novel GCN-based framework for the prediction of stock price movements. Our suggested framework models the relations among stock prices as a novel graph structure called influence network. The algorithm uses a pool of supervised history-based prediction models to label a subset of nodes using the plausible predictions of those models. The partially labeled network is then analyzed by a GCN to extract new representations for the nodes and predict the next direction of price movement for all the stocks. Our experiments showed that the suggested graph-based semi-supervised prediction model significantly outperforms state of the art baseline single-stock and graph-based prediction algorithms. They also show that the introduced algorithm for constructing the influence network as well the initial labeling assignment technique and the final label prediction mechanism, each has a contribution to the high performance achieved by GCNET framework.

The only source of information used by the algorithm was the price history for the stocks. One direction for future research is to consider other sources of information, such as textual information available on the web, to enhance the quality of the graphical modeling of the stocks possible interactions. Another suggestion would be to study the process of assigning initial labels to the nodes, especially by considering the graphical features of the nodes and the amount of initial information that is injected into different parts of the graph.

\bibliography{sample}

\end{document}